\documentclass[12pt]{amsart}
\usepackage{amsthm,amssymb,latexsym}
\usepackage{graphicx, epsfig}
\usepackage{color}

\theoremstyle{plain} 
\newtheorem{thm}{Theorem}
\newtheorem{cor}{Corollary}
\newtheorem{lem}{Lemma}
\newtheorem{prop}{Proposition}

\theoremstyle{definition}

\newtheorem{example}{Example}

\theoremstyle{remark}
\newtheorem{rem}{Remark}

\newcommand{\pp}{\mathbb{P}}

\title[Reconstructing pedigrees]
{Reconstructing pedigrees: A stochastic perspective}

\author{Bhalchandra D. Thatte}

\address{
  Bio\-mathematics Research Centre \\
  Mathematics and Computer Science Building \\
  University of Canter\-bury \\
  Private Bag 4800 \\
  Christ\-church, New Zealand }

\email[Bhalchandra D. Thatte] {bdthatte@gmail.com}

\author{Mike Steel}

\email[Mike Steel] {m.steel@math.canterbury.ac.nz}

\thanks{Supported by the Allan Wilson Centre for Molecular Ecology and
  Evolution}

\date{}

\begin{document}
\begin{abstract}
  A pedigree is a directed graph that describes how individuals are
  related through ancestry in a sexually-reproducing population. In this
  paper we explore the question of whether one can reconstruct a
  pedigree by just observing sequence data for present day
  individuals.  This is motivated by the increasing availa\-bility of
  genomic sequences, but in this paper we take a more theoretical
  approach and consider what models of sequence evolution might
  allow pedigree reconstruction (given sufficiently long sequences).
  Our results complement recent work that showed that pedigree
  reconstruction may be fundamentally impossible if one uses just the
  degrees of relatedness between different extant individuals. We find
  that for certain stochastic processes, pedigrees can be recovered up
  to isomorphism from sufficiently long sequences.
\end{abstract}

\maketitle

\section{Introduction}

Since earliest civilisation people have been concerned with recording,
deciphering and resolving their ancestry. The concept of a `family tree'
is widely familiar (even though the ancestry of an individual cannot
remain a tree for too many generations into the past) and there are many
methods for deciphering ancestry back several generations. Mostly these
are somewhat ad-hoc, based on comparing and combining overlapping
ancestries, oral and written records.

However in recent decades the concept of deeper ancestry has become
topical in molecular evolution. Firstly, the `Out-of-Africa' hypothesis
\cite{cann}, now widely accepted, suggests that all extant humans are
descendants of a relatively small population that migrated (possibly
multiple times) out of Africa around 150,000-200,000 years ago.
Secondly, recent theoretical work \cite{chang} suggests that most of the
human population is likely to have common ancestors much more recently
(thousands rather than hundreds of thousands of years ago). Thirdly,
since the sequen\-cing of the complete human genome in 2001, \cite{gen1,
  gen2} and subsequent improvements in the economics and speed of
sequencing technology, it is quite possible that complete (or
near-complete) genomic sequences for all individuals in a population
could be available in the near future.

These factors immediately suggest the question: what would a very large
amount of genomic data tell us about the ancestry of a popu\-lation?
Clearly one can easily decide who are closely related (siblings, cousins
etc), but how far back in time might one be able to reconstruct an
accurate ancestry? To date, little is known about what is needed in
order to formally reconstruct a pedigree (a graph that describes
ancestry -- defined formally below) though some initial results were
presented in \cite{sh2006}.  This is in marked contrast to another field
in molecular evo\-lution, namely phylogenetics, where there is a
well-developed theory for reconstructing evolutionary (`phylogenetic')
trees on species from the genetic sequences of present-day species
\cite{fels}. In that setting genetic data is often highly informative
for reconstructing detailed relationships between species deep into the
past (tens or hundreds of millions of years). They can also be
informative at short time frames when studying rapidly evolving
organisms (such as HIV).

However in phylogenetics the underlying graph is a tree, while in a
pedigree it is a more `tangled' type of directed graph. Moreover, the
number of vertices in a tree is linearly related to the number of leaves
(which represent the extant species on which we have information) while
for a pedigree the number of vertices (individuals) can keep growing as
we go further back in time.

In this paper we continue the analysis started in \cite{sh2006} and
attempt to determine models under which pedigrees might be reconstructed
from sufficient data. We should point out that there is a well-developed
statistical theory for pedigrees \cite{thom}, but this deals with
different sorts of questions than pedigree reconstruction, such as
estimating an ancestral state in a known pedigree.

In \cite{sh2006} and \cite{thatte15}, pedigrees were considered mainly
from a combinatorial perspective. A question considered in both these
papers was how best to construct pedigrees from certain combinatorial
information about them, such as sets of distances between individuals,
pedigrees on sub-populations, and so on. Several examples and
counterexamples to combinatorial identifiability questions were
presented. It seemed that constructing pedigrees would be a difficult
task, if at all possible, and some of our intuition derived from
phylogenetic trees would not carry over to pedigrees.

A purpose of this paper is to consider pedigrees from a more stochastic
perspective.  We consider several stochastic models of evolution on a
pedigree, that is, mechanisms by which individuals may inherit sequence
information from their parents. We consider the funda\-mental theoretical
question: is the sequence information available in living individuals in
a population sufficient to construct the pedigree of the population, or
might there instead be portions of a pedigree, that will always remain
{\em ghosts}, unable to be clearly resolved regardless of how much
sequence data one has on extant individuals? More formally, we are
interested in whether non-isomorphic pedigrees could produce the same
joint distribution of sequence information for living individuals.  We
begin with some combinatorial preliminaries and enumerate the number of
distinct pedigrees to strengthen an earlier lower bound on the number of
segregating sites that was derived in \cite{sh2006}.

\section{Definitions and preliminaries}

Mostly we follow the notation of \cite{sh2006}. Unless stated otherwise
we will assume all (directed or undirected) graphs are finite, simple
and without loops.  A {\em general pedigree} is a directed acyclic graph
$P = (V,A)$ in which $V$ can be written as the disjoint union of two
subsets $M$ and $F$ (`Male' and `Female'), and where each vertex either
has no-incoming arc or two incoming arcs, with one from a vertex in $M$
and the other from a vertex in $F$.  The vertices with no in-coming arcs
are called the {\em founder vertices}.

In representing ancestry an arc $(u,v)$ of $P$ denotes that $v$ is a
child (offspring) of $u$ (equivalently, $u$ is a parent of $v$), and
the conditions defining a pedigree simply state that each individual
(not in the founding population) has a male and female parent, and that
there is an underlying temporal ordering (acyclicity).

\begin{figure}[ht]
\begin{center}
\includegraphics[width=5in]{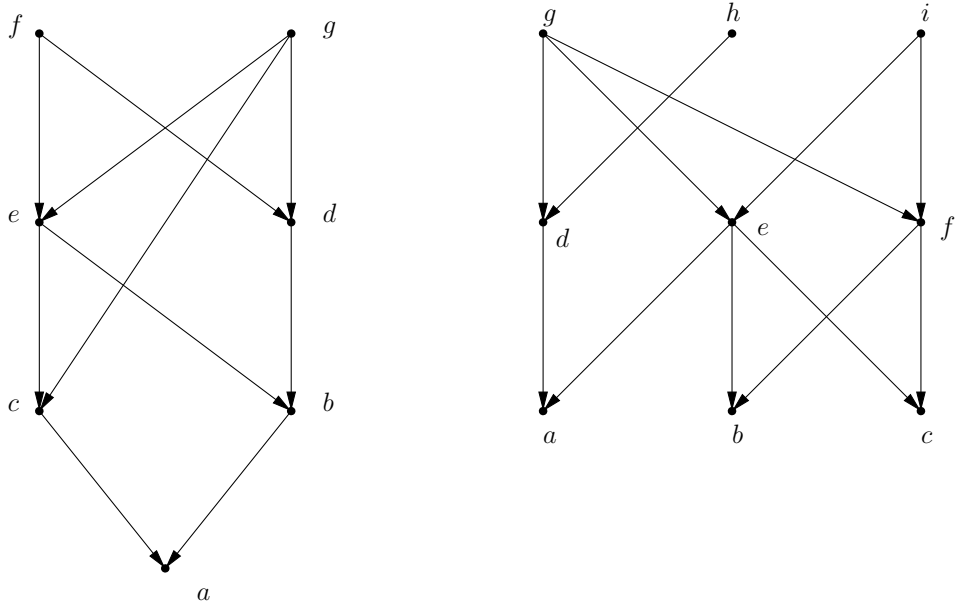}
\end{center}
\caption[]{A general pedigree on $X= \{a\}$ (left) and a simple
pedigree with constant population size on $X = \{a,b,c\}$ (right).}
\label{fig-definition}
\end{figure}

In Figure~\ref{fig-definition}, a general pedigree is shown on the left.

Given a directed graph $G = (V,A)$ let $M(G) = (V, E)$ be the graph on
$V$ whose edge set consists of all pairs $\{u,v\}$ for which there
exists $w \in V$ with $(u,w) \in A$ and $(v,w) \in A$.  In the case
where $G$ is a `food web', $M(G)$ is known as the `competition graph'
(see \cite{mckee}). However in our setting, if $G$ is a pedigree, then
$M(G)$ is the `mate graph' of $G$, where a pair of individuals form an
edge if they have at least one child.

\begin{lem}
  \label{lem1}
  A directed graph $G=(V,A)$ is a pedigree if and only if (i) $G$ is
  acyclic, (ii) $M(G)$ is bipartite, and (iii) no vertex of $G$ has just
  one incoming arc. In particular it can be determined in polynomial
  time (in $|V|$) whether or not a directed graph is a pedigree.
\end{lem}

\begin{proof} Conditions (i)-- (iii) clearly hold if $G$ is a pedigree.
  Conversely, if $M(G)$ is bipartite $V$ can be properly $2$--coloured,
  with colour set $\{M,F\}$, and so we can write $V$ as the disjoint
  union of two sets $M, F$ so that each vertex with at least two
  incoming edges has exactly two incoming edges -- one from a vertex in
  $M$ and one from a vertex in $F$.  Condition (iii) excludes the
  possibility of just one incoming edge, and so $G$ is a pedigree. For
  the second claim, observe that the three conditions (i)--(iii) can all
  be established in polynomial time.
\end{proof}

The set of vertices that have no out-going arcs is denoted $X_0$, and
for a particular distinguished subset $X$ of $X_0$ (called the {\em
  extant individuals}) we refer to $(P,A)$ as a {\em pedigree on $X$}.
We assume that the vertices in $X$ are labelled, and other vertices are
unlabelled.  Two pedigrees on $X$ are {\em isomorphic} if there is a
diagraph isomorphism between them that fixes each element of $X$.

We note in passing that in \cite{sh2006} it was sometimes assumed that
the decomposition $(M, F)$ of $V$ was known, as this is not necessarily
uniquely determined just by $P$; this in turn also allows a more
restrictive defi\-nition of isomorphism (called `gender-isomorphism') in
which the diagraph isomorphism is required to map $M$ (resp. $F$)
vertices to $M$ (resp. $F$) vertices.  However we do not require or
invoke this additional structure in the current paper.

A {\em simple pedigree} is a pedigree in which the vertex set of the
pedigree is a disjoint union of $X_i; 0 \leq i \leq d$, and every arc
$(u,v)$ has its tail $u$ in $X_i$ and its head $v$ in $X_{i-1}$, for
some $i > 0$. In this case, $X_0$ is the set of extant vertices, and
$X_d$ is the set of founders, and $d$ is the depth of the pedigree.  In
\cite{sh2006} and \cite{thatte15}, the term `discrete generation
pedigree' was used instead of the term `simple pedigree'. In simple
pedigrees with a {\em constant population size}, all $X_i$ have the same
cardinality. In Figure~\ref{fig-definition}, a simple pedigree with a
constant population size is shown on the right.

The amount of information required to accurately reconstruct a pedigrees
on a set of size $n$, and up to depth $d$ is clearly bounded below by
some increasing function of the number of distinct (mutually
non-isomorphic) simple pedigrees with a constant population size $n$ and
of depth $d$.  Let this number be $f(n,d)$.  We first describe a lower
bound on $f(n,d)$ providing a slightly stronger bound than
\cite{sh2006}.

Let $X_0 = \{x_i; 1 \leq i \leq n\}$ and $X_1 = \{y_i; 1 \leq i \leq
n\}$. Consider a tree $T$ defined on $X_1$. We construct a pedigree on
$X_0\cup X_1$ with the set of extant vertices $X_0$ as follows: we first
take an arbitrary onto map $g$ from $X_0$ to the edge set $E(T)$ of $T$,
and for every $x_k \in X_0$, if $g(x_k) = \{y_i,y_j\}$, then in the
pedigree, $x_k$ is a child of $y_i$ and $y_j$. We count the number of
pedigrees that can be constructed in this manner by considering all
possible mutually non-isomorphic trees $T$, and all possible onto maps
from $X_0$ to $E(T)$. For a fixed tree $T$, there are exactly
$\binom{n}{2}(n-1)!$ onto maps from $X_0$ to $E(T)$. Each map does not
give us a distinct pedigree; in fact, each pedigree constructed this way
is repeated $|\text{aut} T|$ times, where $\text{aut} T$ is the
automorphism group of $T$.  Thus we have

\begin{displaymath}
  f(n,1) \geq \sum_{T}\frac{\binom{n}{2}(n-1)!}{|\text{aut} T|},
\end{displaymath}
where the summation is over all mutually non-isomorphic trees on $X_1$.
Since $n!/|\text{aut} T|$ is the number of labelled trees isomorphic to
a given tree $T$, summing over all mutually non-isomorphic trees gives
us

\begin{displaymath}
  f(n,1) \geq \frac{(n-1)n^{n-2}}{2},
\end{displaymath}
where $n^{n-2}$ is the number of labelled trees on $X_1$, by Cayley's
classic formula \cite{cay}.

Observe that each vertex in $X_1$ is {\em distinguished} in the
pedigree, in the sense that no two vertices in $X_1$ have the same set
of children.  This fact is useful to construct distinct pedigrees of
arbitrary depth by repeating the same construction for arcs between
$X_1$ and $X_2$, $X_2$ and $X_3, \ldots, $ Therefore,

\begin{displaymath}
  f(n,d) \geq \frac{(n-1)^dn^{d(n-2)}}{2^d}
\end{displaymath}

Observe also that, since trees are bipartite, the directed graph
constructed is indeed a pedigree by Lemma~\ref{lem1}.

The above estimate gives an information theoretic lower bound of
$(d/2)\log n +o(\log n)$ on the number of segregating sites needed for
reconstructing a pedigree from DNA sequence data. This follows by the
same argument as in \cite{sh2006} and is a slight improvement on the
bound $(d/3)\log n + o(\log n)$ established in that paper.

\section{Pedigree reconstruction}
\label{sec-stochastic}

In this section, we examine the question of constructing a pedigree from
the information obtained from the extant individuals. In biological
applications, this information is typically provided by (DNA) sequence
data. It is assumed that the information has been passed on to each
individual by its parents; and, over generations, the information
undergoes a stochastic change that models the evolutionary process. Is
the information available at all extant individuals sufficient to
uniquely construct the pedigree of the population? To be precise, are
there examples of stochastic processes for which we cannot construct the
pedigree, and are there examples of stochastic processes for which we
can construct the pedigree?

\subsection{A negative result}

We begin with a simple Markov process under which the information at the
extant vertices (in the form of binary sequences of arbitrary length) is
not sufficient to uniquely determine the pedigree.

Suppose $\{u_i; 1 \leq i \leq p\}$ is the vertex set of a pedigree
$\mathcal{P}$. Suppose that associated with each vertex $u_i$ in the
pedigree $\mathcal{P}$, there is a random variable $U_i$ that takes
values from a finite state space $S$. Let $$\mathbb{P}(U_i = a_i| U_j =
a_j; 1\leq j \leq p, j \neq i)$$ denote the probability that $U_i$ takes
the value $a_i$ conditional on the states of random variables at all
other vertices. We assume that $$\mathbb{P}(U_i = a_i| U_j = a_j; 1 \leq
j \leq p, j \neq i) = \mathbb{P}(U_i = a_i| U_j = a_j, U_k = a_k),$$
where $u_j$ and $u_k$ are the parents of $u_i$. Is it possible to
construct the pedigree up to isomorphism given the joint distribution
$\mathbb{P}(U_1 = a_1, U_2 = a_2, \ldots, U_n = a_n)$, where we use the
indices 1 to $n$ for extant vertices?

Consider a symmetric two-state model given by the transition matrix

\begin{center}
\begin{tabular}{r|c|c|c|c}
   & 00          & 01   & 10   & 11         \\ \hline
0  & $\alpha$    & 0.5  & 0.5  & $1-\alpha$ \\ \hline
1  & $1-\alpha$  & 0.5  & 0.5  & $\alpha  $ \\ \hline
\end{tabular}
\end{center}
where the columns are indexed by the joint states of the parents of a
vertex, and the rows are indexed by the state of the vertex. For
example, the entry in the first column and second row says that the
probability that a child is in state 1 conditional on both parents being
in state 0 is $1 - \alpha $.

In the following, we construct non-isomorphic pedigrees $\mathcal{P}$
and $\mathcal{Q}$, each on two extant vertices $u_1$ and $u_2$, such
that the joint distribution $\mathbb{P}(U_1=a_1, U_2=a_2)$, where $a_i
\in \{0,1\}$, is identical for $\mathcal{P}$ and $\mathcal{Q}$.

\begin{enumerate}

\item Construct two disjoint binary pedigrees $\mathcal{B}_i; i \in\{
  1,2\}$, respectively, on extant vertices $u_1$ and $u_2$. The depth of
  each binary pedigree is $t \geq 2$. Let $S_i; i \in \{1,2\}$ be the
  corresponding sets of their founders.

\item Construct a single intermediate pedigree $\mathcal{P}^\prime$ from
  $\mathcal{B}_i; i \in\{1,2\}$ by identifying each vertex in $S_1$ with
  a unique vertex in $S_2$. Construct pedigree $\mathcal{P}$ by adding
  vertices $v$ and $w$ as parents of all founder vertices in the
  pedigree $\mathcal{P}^\prime$.

\item Construct pedigree $\mathcal{Q}$ as in the above step so that
  $\mathcal{P}$ and $\mathcal{Q}$ are non-isomorphic. This is possible
  when $t \geq 2$.

\end{enumerate}

\begin{figure}[ht]
\begin{center}
\includegraphics[width=5in]{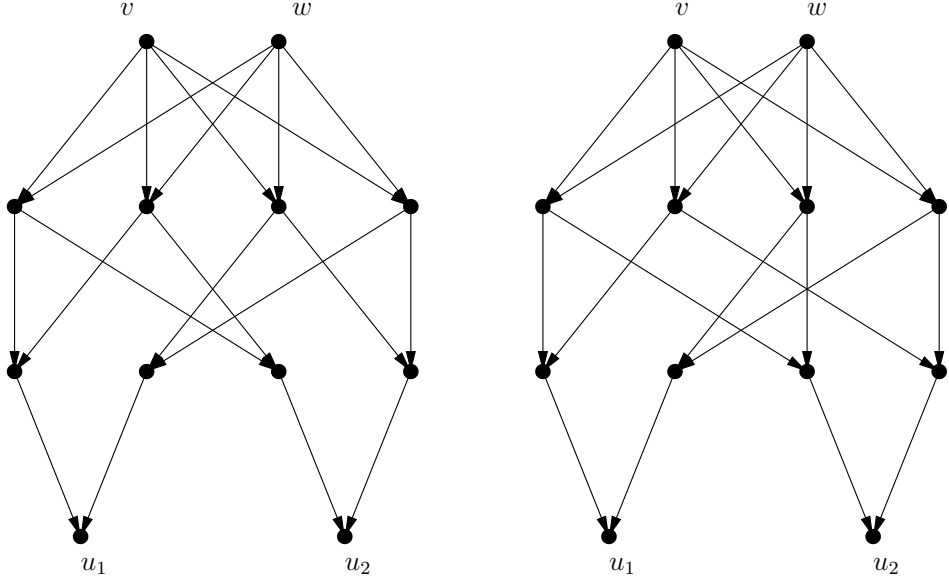}
\end{center}
\caption[]{Non-isomorphic pedigrees that produce indistinguishable
sequences under the symmetric
  stochastic model.}
\label{fig-stochastic}
\end{figure}

Figure~\ref{fig-stochastic} shows examples of $\mathcal{P}$ and
$\mathcal{Q}$ for $t = 2$.

\begin{prop} The pedigrees $\mathcal{P}$ and $\mathcal{Q}$ have the same
  joint distribution $\mathbb{P}(U_1=a_1, U_2=a_2)$, where $a_i \in
  \{0,1\}$, under the symmetric model described above. Thus the two
  pedigrees cannot be distinguished from each other from binary
  sequences (of i.i.d.  samples) of any finite (or infinite) length.
  \end{prop}

\begin{proof}

  First consider a binary pedigree, say $\mathcal{B}_1$. Let $k$ of the
  vertices in $S_1$ be in state 0. Let $f(k,t)$ denote the probability
  that the vertex $u_1$ is in state 0.  Suppose $k_1$ of the 0 states
  occur among the founders on the left tree, and $k_2$ occur on the
  right tree, where the left tree and the right tree are the pedigrees
  of the two parents of $u_1$. Therefore, $k_1 + k_2 = k$.  A recurrence
  for $f(k,t)$ is then written in terms of $f_1 = f(k_1,t-1)$ and $f_2 =
  f(k_2,t-1)$.
  \[
  f(k,t) = \alpha f_1f_2 + 0.5(1-f_1)f_2 +
  0.5f_1(1-f_2)+(1-\alpha)(1-f_1)(1-f_2),
  \]
  where the four terms correspond to the four possible joint states of
  the parents of $u_1$.

  It can be verified by induction that the following expression for
  $f(k,t)$ solves the recurrence.
  \[
  f(k,t) = \frac{k}{2^t}(2\alpha -1)^t + \frac{1-(2\alpha -1)^t}{2}.
  \]
  Here the independence of $f(k,t)$ on exactly where the zero states
  occur among the founders is what is useful in the following.

  Now consider the intermediate pedigree $\mathcal{P}^\prime$ and
  consider the event $E_k$ that exactly $k$ of its founders are in state
  0 (so $k \in \{0,1,2,3,4\}$). The conditional probability
  $\mathbb{P}(U_1 = a_1, U_2 = a_2|E_k)$ is given by \[ \mathbb{P}(U_1 =
  a_1, U_2 = a_2|E_k) = \mathbb{P}(U_1 = a_1|E_k)\mathbb{P}(U_2 =
  a_2|E_k), \] where each factors is either $f(k,t)$ or $1-f(k,t)$
  depending on whether $a_i$ are 0 or 1, respectively. This is also true
  in $\mathcal{Q}^\prime$.

  The vertices $v$ and $w$ are added to both intermediate pedigrees as
  parents of vertices in $S_1$ and $S_2$ so as to guarantee that all
  possible joint states on $S_i$ that have $k$ zeros are equally likely.
  This implies that for any given joint distribution on $v$ and $w$, we
  have the same joint distribution on $u_1$ and $u_2$ in $\mathcal{P}$
  and $\mathcal{Q}$.
\end{proof}

We now show that exponentially many mutually non-isomorphic pedigrees
can be obtained by this construction.

\begin{prop}
  The number of mutually non-isomorphic pedigrees that can be obtained
  by the above construction grows super-exponentially with $t$.
\end{prop}

\begin{proof}

  Consider two disjoint binary pedigrees $\mathcal{B}_i$ of depth $t
  \geq 2$, on extant vertices $u_i$, and founder sets $S_i$, where $i
  \in\{ 1,2\}$. Let $|S_i| = 2^t = m$. There are $m!$ ways of
  identifying vertices in $S_2$ with vertices in $S_1$, but not all of
  them result in mutually non-isomorphic pedigrees. Consider a pedigree
  $\mathcal{P}^\prime$ obtained by identifying vertices in $S_2$ with
  vertices in $S_1$. The automorphism group of $\mathcal{P}^\prime$ is a
  subgroup of the automorphism group of $B_1$. But $|\text{aut}\, B_1|$
  is $2^{m-1}$, therefore, $|\text{aut}\, \mathcal{P}^\prime| \leq
  2^{m-1}$.  Therefore, the number of mutually non-isomorphic pedigrees
  obtained by identifying vertices of $S_2$ with vertices in $S_1$ is at
  least
  \[
  \frac{m!}{2^{m-1}},
  \]
  which implies the claim.
\end{proof}

\subsection{Positive results}

We first describe a simple deterministic process, and a related
stochastic variation, under which the information available at the
extant individuals is sufficient to construct the pedigree.  We then
describe a Markov model that comes closer to the mutation-recombination
setting of genetic ancestry, for which pedigree reconstruction is also
possible. This last model should be viewed as a proof-of-concept, rather
than as realistic processes that capture all aspects of evolu\-tionary
processes.

\begin{example}[Deterministic process] \label{ex-deterministic} Suppose
  each founder in the popu\-lation has a distinct label. Consider an
  individual whose parents are labelled $Y$ and $Z$. Suppose that each
  individual inherits the labels of its parents, and also has its own
  unique character that is not seen before in any other individual. In
  this way we assign the individual a label $\{\{Y,Z\}, X\}$, where $X$
  is a new symbol or a trait that no other individuals in the
  population, except for descendents of the individual under
  consideration, who inherit $X$ in the manner described.

  From the labels of the extant individuals, the pedigree is uniquely
  constructed in a straight forward manner. First we construct the
  pedigree of each extant individual. Each individual's label uniquely
  determines the labels of its parents and the new character that has
  arisen in the population for the first time. We recursively construct
  a binary tree of parents, grand parents, ... beginning with an extant
  individual. After constructing the binary tree, we identify vertices
  that have the same labels. Such vertices are ancestors to whom there
  are multiple paths from the extant individual.

  The next step is to construct a (graph theoretic) union of pedigrees
  of all extant individuals. In constructing the graph theoretic union,
  vertices in different pedigrees that have the same labels are
  identified, and multiple arcs between two vertices are suppressed to
  leave a single arc between them. This completes the construction.
  \end{example}

\begin{example}[Semi-deterministic process on the integers]
  \label{ex-random1} Now we modify Example ~\ref{ex-deterministic} so as
  to introduce some randomness, and also to work over a fixed state
  space (the integers).  Let $N$ be a large positive integer
  (sufficiently large relative to the number of vertices in the
  pedigree, in a sense that will be made more precise shortly). To each
  individual $i$ in the pedigree we first associate an independent
  random variable $Y_i$ that takes a value selected uniformly at random
  from $\{1,\ldots, N\}$. We then assign a random state $X_i$ to each
  vertex $i$ of the pedigree as follows. If $i$ is a founder, then set
  $X_i = Y_i$. Otherwise, if $i$ has parents $j$ and $k$ then set

  \begin{displaymath}
    X_i = 2^{X_j+N} + 2^{X_k+N} + Y_i.
  \end{displaymath}

  Observe that this process is Markovian (the state at a vertex depends
  just on the states at the parents, and not on earlier ancestors).
  Moreover, if the random variables $Y_i$ take distinct values, then the
  pedigree can be uniquely constructed since $2^{a+N} + 2^{b+N} + m$ can
  be uniquely `decoded' as $\{\{a,b\},m\}$. If there are $n$ vertices
  in the pedigree (and $N \geq n$) the probability that each random
  variable takes a distinct value is

  \begin{displaymath}
    \frac{N(N-1)\ldots (N-n+1)}{N^n},
  \end{displaymath}
  which approaches 1 as $N$ tends to infinity.

  Therefore, under this process, a pedigree can be uniquely
  reconstructed by observing the random variables at the extant
  vertices, with a probability approaching 1 as $N$ tends to infinity.
\end{example}

Although the above examples seems to be far removed from the reality of
biological evolution, the concept underlying the examples is almost
un-recognisably hidden in the following setting where the main
consideration is to construct a process that models sequence evolution.

\section{A stochastic process on sequences that allows reconstruction}
\label{ex-random2}

The process of inheriting genetic material from parents may be
conceptualised as follows. Suppose the parents $Y$ and $Z$ of an
individual $X$ have sequences $\{y_i; i = 1, 2, \ldots \}$ and $\{z_i; i
= 1, 2, \ldots \}$, respectively. Here the sequences are assumed to be
sequences of characters drawn from $[N] = \{1,2,\ldots , N\}$. We assume
that the sequence $\{x_i\}$ of $X$ is constructed by copying segments of
sequences $\{y_i\}$ and $\{z_i\}$ so that roughly half the genetic
material is inherited from one parent, and roughly half from the other
parent. In addition to the directly copied bits and pieces from its
parents' genetic sequences, $X$ also has in its sequence occurrences of
segments that are not (recognised as) copies of segments of $\{y_i\}$
and $\{z_i\}$.  We suppose that the $X$-specific fragments are
constructed from characters drawn from a set $U_X\subset [N]; |U_X| =
m$, where $U_X$ is chosen uniformly at random from the family of all
subsets of $[N]$ of cardinality $m$. The process of construction of the
sequence $\{x_i; i = 1, 2, \ldots\}$ is then modelled as in a hidden
Markov model.  The copying process copies character from $\{y_i\}$, and
at some step, determined by chance, begins copying characters from
$\{z_i\}$, or begins a random generation of a sequence of characters
chosen from $U_X$. The process of copying from and switching between
$\{y_i\}$, $\{z_i\}$ and $U_X$ continues.

But the segments copied from $\{y_i\}$ and $\{z_i\}$ are in turn partly
inherited from the parents of $Y$ and $Z$, respectively, and partly from
the $Y$-specific and $Z$-specific segments, that is, segments of
characters drawn from $U_Y$ and $U_Z$, respectively.

We model the above description by first defining a one to one
corres\-pondence between pedigrees and a subclass of finite automata
that {\em emit} (to use the HMM terminology) character sequences at the
extant individuals. We then demonstrate how a sufficiently long emitted
sequence determines first the automaton and then the pedigree with high
probability.

Without a loss of generality, we consider pedigrees with a single extant
vertex, since after constructing all sub-pedigrees having a single
extant vertex, we can construct their graph theoretic union, as in
Example ~\ref{ex-deterministic}. This is discussed further in Remark
~\ref{rem-union}.

\subsection{The automaton (directed graph) $G$, and the mechanism of
sequence emission.} \label{sec-auto}

Let $\mathcal{Q}$ be a pedigree with vertex set $V; |V| = n$, with a
single extant vertex $x$. The automaton associated with $\mathcal{Q}$ is
denoted by a directed graph $G$ on the vertex set $V$. For convenience,
we have used the same vertex set $V$; so to avoid ambiguity, we denote
an arc from $y$ to $z$ in $\mathcal{Q}$ by $yz$, and an arc from $y$ to
$z$ in $G$ by $(y,z)$.

The automaton $G$, its transition probabilities, and the mechanism by
which it emits characters in the sequence of the extant vertex are
defined so that the following conditions are satisfied.

\begin{enumerate}

\item Let $[\delta_1, \delta_2] \subseteq [0,1]$ and $[\Delta_1,
\Delta_2]\subseteq [0,1]$ be two intervals such that $\delta_i$ are
much smaller that $\Delta_j$ for $i,j \in \{1,2\}$.

\item For each internal vertex $y$, (that is, a vertex that is neither a
founder vertex nor the extant vertex), there are two arcs $(y,u)$ and
$(y,v)$ to its parents $u$ and $v$, respectively, an arc $(y,x)$ to the
extant vertex $x$, and a self loop. We assume that the transition
probabilities satisfy \begin{displaymath} p(y,u), p(y,v) \in [\Delta_1,
\Delta_2] \end{displaymath} and \begin{displaymath} p(y,x), p(y,y) \in
[\delta_1, \delta_2]. \end{displaymath}

\item For the extant vertex $x$, there are outgoing arcs $(x,y)$ and
$(x,z)$ to its parents, $y$ and $z$, respectively, and a self-loop, with
the corresponding transition probabilities given by \begin{displaymath}
p(x,y), p(x,z) \in [\Delta_1, \Delta_2] \end{displaymath} and
\begin{displaymath} p(x,x)+p(x,y)+p(x,z) = 1. \end{displaymath}

\item From a founder vertex $z$, there is one arc $(z,x)$ to the extant
vertex $x$, and a self-loop. The transition probabilities satisfy
\begin{displaymath} \delta_1 \leq p(z,x) \leq \delta_2 \end{displaymath}
and \begin{displaymath} p(z,x)+p(z,z) = 1. \end{displaymath}

\item Each vertex $y$ of the automaton corresponds to a subset $U_y$ of
$[N]$, such that $|U_y| = m > 1$, and $U_y$ is chosen randomly from a
uniform distri\-bution on the family of subsets of $[N]$ of cardinality
$m$. The character sequence for $x$ is emitted by the automaton as
follows: the automaton defines a Markov chain with transition
probabilities defined above; when the chain is in state $y$, (that is,
at vertex $y$ of the auto\-maton), a character from $U_y$ is emitted
from a uniform distri\-bution on $U_y; y \in V$.

\end{enumerate}

The assumption that $\delta_i$ are much smaller than $\Delta_j$ for $i,j
\in \{1,2\}$, and the conditions listed above imply that an individual
derives most of its genetic material from its parents, who in turn
receive most of their genetic material from their parents.

\begin{figure}[ht]
\begin{center}
\includegraphics[width=5in]{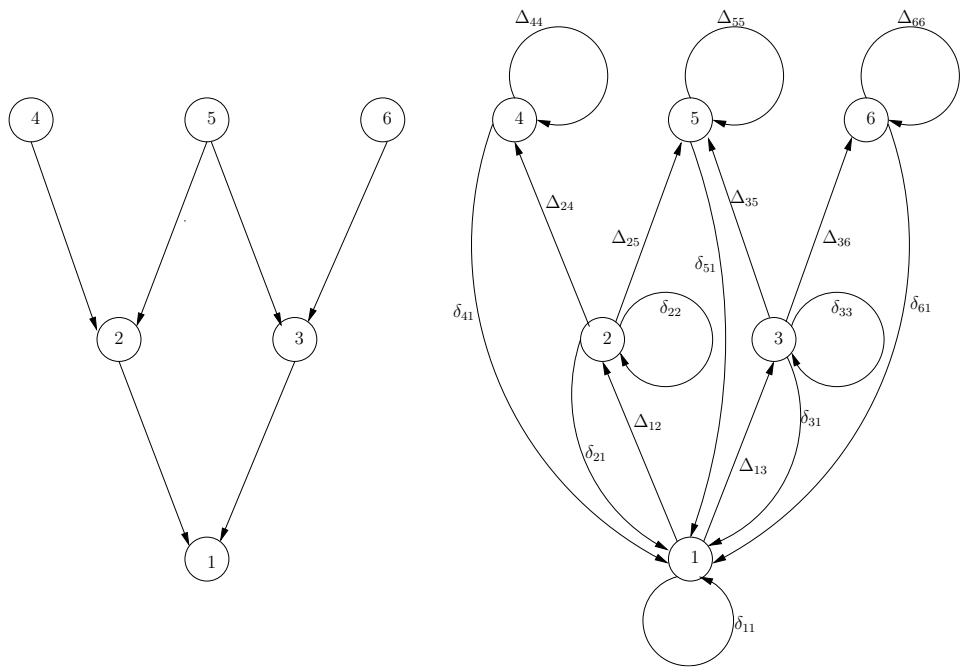}
\end{center}
\caption[]{A pedigree and a corresponding automaton.}
\label{fig-automaton}
\end{figure}

Figure~\ref{fig-automaton} shows a pedigree $\mathcal{Q}$ on 6 vertices
and an automaton $G$ that corresponds to the pedigree $\mathcal{Q}$. The
transition probabilities in the figure are denoted by $\Delta_{ij}$ or
$\delta_{ij}$ instead of $p(i,j)$ so as to indicate their relative
magnitudes.

We are interested in the following question: does a sufficiently
long sequence $\{x_i; i = 1, \ldots \}$ emitted by the automaton
determine the pedigree unambiguously with high probability?  Since
the correspondence between the subclass of automata and pedigrees
with a single extant vertex is one-to-one, the question is
equivalent to asking if the auto\-maton can be constructed
unambiguously. The main result of this section is the affirmative
answer to this question, formulated in the following theorem. Note
that although it deals with only a single extant vertex, we describe
in Remark~\ref{rem-union} how it extends to the general case of a
pedigree over a finite set $X$.

\begin{thm} \label{thm-main} Let $\mathcal{Q}$ be a pedigree having a
single extant vertex. Let $\mathcal{Q}$ be associated with an
automaton $G$ that satisfies the conditions listed above. Let $S_k =
\{x_i; i = 1,2,\ldots, k\}$ be a sequence of characters from the set
$[N] = \{1,2,\ldots, N\}$, emitted by the automaton (as in the fifth
condition above). Then for sufficiently large $k$ and $N$, the
automaton $G$ and the pedigree $\mathcal{Q}$ can be correctly
reconstructed (with high probability) from the sequence $S_k$.
\end{thm}

The theorem follows from the several lemmas proved next.

\begin{lem} Given an automaton $G$ with its transition probabilities,
the pedigree $\mathcal{Q}$ can be uniquely constructed. \end{lem}

\begin{proof} This follows from the relative magnitudes of the
probabilities of transition. For distinct vertices $u$ and $v$ in $G$,
the transition probability from $u$ to $v$ is {\em high}, (that is, in
the interval $[\Delta_1,\Delta_2]$), if and only if $v$ is a parent of
$u$ in the pedigree $\mathcal{Q}$. For a vertex $u$, the probability of
transition from $u$ to itself is {\em high} if and only if $u$ is a
founder vertex. A vertex $u$ is the extant vertex of $Q$ if and only if
there is no other vertex $v$ in $G$ such that the probability of
transition from $v$ to $u$ is {\em high}.  \end{proof}

Next we must construct the automaton $G$ from the sequence $S_k$. The
idea of inference of the automaton $G$ from the sequence $S_k$ is based
on the following observation. Suppose $i,j \in [N]$ are such that there
is only one $U_y$ that contains $i$, and only one $U_z$ that contains
$j$. Then the observed transition probability $p(i|j)$ in the sequence
$S_k$ is in the range $[\Delta_1/ m, \Delta_2/m]$ if $y$ is a parent of
$z$; and is in the range $[\delta_1/ m, \delta_2/m]$ if $i\in U_x$ and
$j\in U_y$, or if $\{i,j\}\subseteq U_y$, where $y$ is an internal
vertex. Similarly, one can argue about the magnitude of the observed
frequency of $i$ followed by $j$ in $S_k$ for founder vertices, and for
the extant vertex. What matters is whether the estimated probability is
{\em high} (of the order of $\Delta_i/m; i = 1,2$) or {\em low} (of the
order of $\delta_i/m$; i = 1,2). The transition probabilities $p(i|j)$
can be estimated as accurately as desired by choosing sufficiently large
$k$. It is crucial for the above argument that each $U_y$ contains some
state $i$ that is unique to $U_y$, that is, $i$ does not belong to a
$U_z$ for $z$ other than $y$. This is the case with high probability for
large $N$, as made precise in the following lemma.

\begin{lem}
Suppose that the sets $U_y$ are randomly chosen from a uniform
distribution on the family of subsets of $[N]$ of cardinality $m$.  Let
$E$ be the event that each $U_y$ contains at least one $i$ that is not
in any other $U_z$. The probability of this event $E$ approaches 1 as
$N$ tends to infinity.
\end{lem}

\begin{proof} Let $E_i$ be the event that $U_i$ is not a subset of
$\cup_{j \neq i} U_j$.  Then, $E = \cap_{i=1}^n E_i$, and by Boole's
inequality \cite{grim}, and symmetry,
\begin{displaymath}
\pp(E) \geq 1- \sum_{i=1}^n \pp(E^c) = 1- n\pp(E_1^c),
\end{displaymath}
where the superscript $c$ denotes complement. Now $E_1^c$ is the
event that $U_1$ is a subset of $U_2 \cup E_3 \cup \ldots \cup U_n$,
and clearly the probability of this (complementary) event is
maximised if $U_2,\ldots, U_n$ are disjoint.  In this case $|U_2
\cup....U_n| = (n-1)m$, and so $\pp(E_1^c)$ is bounded above by the
proportion of subsets of $[N]$ of size $m$ that are subsets of a set
of size $(n-1)m$, i.e. $\pp(E_1^c) \leq
\frac{\binom{(n-1)m}{m}}{\binom{N}{m}}$. This, along with the above
inequality, implies $\pp(E) \rightarrow 1$ as $N \rightarrow \infty
$.
\end{proof}

Let $U_i\subseteq [N]; i = 1,2,\ldots n$ be the unknown character sets
corresponding to the vertices $1,2, \ldots , n$ of the automaton. Let
$\bar{U}_i$ denote the subset of $U_i$ consisting of those elements that
are unique to $U_i$, that is,
\[
\bar{U}_i = U_i \cap (\cup_{j\neq i}U_j)^c.
\]

By a recursive procedure, we construct $\bar{U}_i$, and the pedigree
$\mathcal{Q}$ on the vertex set $[n] = \{1,2,\ldots , n\}$.

Without a loss of generality, assume that the extant vertex is labelled
1, and the founder vertices are labelled from $f$ to $n$.

We first construct a directed graph $H$ from the observed sequence $x_i;
i = 1,2,\ldots $. The vertex set $V(H)$ of $H$ is the set of states that
appear in the emitted sequence $x_i; i = 1,2,\ldots$. The set of arcs of
$H$ is $E(H)$, and an arc $(u,v)$ is in $E(H)$ if a transition from $u$
to $v$ is observed in $x_i; i = 1,2,\ldots $, that is, if there is some
$i$ for which $x_i = u$ and $x_{i+1} = v$. Each arc $(u,v)$ of $H$ is
labelled {\em high} or {\em low} depending on whether the inferred
probability $p(v|u)$ of transition from $u$ to $v$ is of the order of
$\Delta/m$ or $\delta /m$, respectively, where $\Delta_1 \leq \Delta
\leq \Delta_2$ and $\delta_1 \leq \delta \leq \delta_2$. The inferred
probabilities will be distinguishable as {\em high } or {\em low} for
sufficiently long emitted sequences.

Let $d^+_h(u)$ and $d^+_l(u)$ denote the number of outgoing arcs from
$u$ that are labelled {\em high} and {\em low}, respectively. We count
each self-loop as a single arc.

\begin{lem} The sets $\bar{U}_i$ and $U_i$ for founder vertices can be
constructed.  \end{lem}

\begin{proof} Suppose $i$ is a founder vertex. Then from a state $u$ in
$\bar{U_i}$, there are precisely $m$ transitions with {\em high}
probability. On the other hand, if $i$ is not a founder vertex, then it
has parents $j$ and $k$; therefore, from a state $u$ in $U_i$, there are
at least $|U_j \cup U_k| \geq m+1$ outgoing arcs that are labelled {\em
high}.  Observe also that if $i$ a founder vertex, and $u$ is in $U_i$
but not in $\bar{U}_i$ then there will be at least $m+1$ outgoing arcs
from $u$ that are labelled {\em high}, since $u$ will also be in some
other $U_j$ in that case. Therefore, $u$ is in $\bar{U}_i$ for some
founder vertex $i$ if and only if $d^+_h(u)=m$. The set of all such
vertices in $H$ naturally partitions into blocks, one block $\bar{U}_i$
for each founder $i$, since if $\bar{U}_i$ and $\bar{U}_j$ correspond to
two founders, and $u \in \bar{U}_i$ and $v \in \bar{U}_j$ then there
will be transitions from $u$ to $v$ and from $v$ to $u$ in the emitted
sequence if and only if $\bar{U}_i = \bar{U}_j$. Once $\bar{U}_i$ is
known for each founder $i$, we can construct $U_i$ as well: if there is
an arc $(u,v)$ that is labelled {\em high} for a state $u$ in
$\bar{U}_i$ and a state $v$ not in $\bar{U}_i$, where $i$ is a founder
vertex, then $v$ must be in $U_i$.  \end{proof}

In general, for vertices other than founders, we will be interested
in constructing only $\bar{U}_i$.

We treat the above construction as the base case of a recursive
procedure for constructing all $\bar{U}_i$.

Let $\mathcal{F} = \{\bar{U}_i\}$ be the collection that has been
constructed so far. At the end of the base case, each $\bar{U}_i; i \geq
f$ is in $\mathcal{F}$.  The construction proceeds in a {\em top-down}
manner; so if $j$ and $k$ are parents of $i$, and if $\bar{U}_i$ is in
$\mathcal{F}$, then $\bar{U}_j$ and $\bar{U}_k$ have already been
constructed and added in $\mathcal{F}$. Let $\cup_S$ denote the union
over all sets in $\mathcal{F}$.

Let $\bar{U}_j$ and $\bar{U}_k$ be any two distinct sets in
$\mathcal{F}$ such that $\bar{U}_i$ for children $i$ with parents $j$
and $k$ have not been constructed so far.

Let $T_{jk}$ be the set of states $u$ for which the following conditions
hold:

\begin{enumerate}

\item $u$ is not in $\cup_S\cup_{r\geq f} U_r$, and

\item there is a {\em high} arc $(u,w)$ in $H$ for every $w$ in
$\bar{U}_j \cup \bar{U}_k$

\end{enumerate}

\begin{lem} If a state $u$ is in $T_{jk}$ then it is in $U_i$ for some
child $i$ with parents $j$ and $k$. If a state $u$ is in $\bar{U}_i$ for
some child $i$ with parents $j$ and $k$ then $u$ is in $T_{jk}$.

\end{lem}

\begin{proof} When the second condition holds it is possible that $u$ is
in $U_j\cap U_k$ and both $j$ and $k$ are founders. But this possibility
is eliminated by the first condition. Therefore $u$ must be in $U_i$ for
some child $i$ with parents $j$ and $k$. The second statement is then
obvious.  \end{proof}

The above proposition implies that \[ \cup_i \bar{U}_i \subseteq T_{jk}
\subseteq \cup_i U_i, \] where the unions are over the children of $j$
and $k$.

\begin{lem} Let $u$ be a state in $T_{jk}$. If $u$ is in $\bar{U}_i$ for
some child $i$ with parents $j$ and $k$ then $d^+_h(u) = |U_j\cup U_k|$,
(which may not be known). If $u$ is not in $\bar{U}_i$ for any child $i$
with parents $j$ and $k$, then $d^+_h(u) \geq |U_j\cup U_k| + 1$
\end{lem}

\begin{proof} The first statement follows from the fact that $u$ is not
in any other set $U_r$, and the second statement follows from the fact
that $u$ is in $U_i$ for some child $i$ with parents $j$ and $k$ and at
least in one other $U_r$.  \end{proof}

\begin{cor} The set $\bar{T}_{jk} = \cup_i \bar{U}_i$, where the union
is over children $i$ of $j$ and $k$, is recognised.  \end{cor}

\begin{proof} The set $\bar{T}_{jk}$ is the set of states $u$ in
$T_{jk}$ for which $d^+_h(u)$ is minimum.  \end{proof}

\begin{lem} The set $\bar{T}_{jk}$ partitions into blocks $\bar{U}_i$
for children $i$ with parents $j$ and $k$.
\end{lem}

\begin{proof} States $v$ and $w$ in $\bar{T}_{jk}$ are the same block if
and only if there are arcs $(v,w)$ and $(w,v)$ labelled {\em low}.
\end{proof}

This construction terminates when no more blocks can be added to
$\mathcal{F}$, thus completing the proof of Theorem~\ref{thm-main}.

\begin{rem} \label{rem-union}

In the above construction we recognised $\bar{U}_i$ for all vertices
in the pedigree. We also recognised the parent-child relationships
between them, which allowed us to construct the whole pedigree on
the single extant vertex. Now suppose that we have a pedigree on
more than one extant individuals. For each extant vertex we have a
sequence emitted by the automaton that corresponds to the
sub-pedigree on that extant vertex. It is reasonable to suppose that
each vertex $i$ in the pedigree corresponds to a unique $U_i \subset
[N]$. Such a supposition means that the extant individuals that are
descendents of $i$ (the {\em cluster} of $i$) share some common
traits, and the states in $\bar{U}_i$ are observed only in the
sequences of the extant individuals in the cluster of $i$. We,
therefore, construct the pedigree of each extant individual
separately. To construct a graph theoretic union of all these
pedigrees, we identify vertices $y$ and $z$, respectively, in
pedigrees $\mathcal{P}_i$ and $\mathcal{P}_j$ whenever $\bar{U}_y$
and $\bar{U}_z$ are identical. It is possible to generalise the
correspondence between pedigrees and automata that was considered
above to a correspondence between pedigrees on multiple extant
vertices and more general automata in which there are transitions
from a vertex either to its parents or to itself or to any of its
extant descendents. The mechanism for emitting characters would not
be essentially different. For example, when the automaton is in
state $v$, (that is, at vertex $v$), it would emit characters from
$U_v$ at all its descendents.  \end{rem}
\subsection{Example} We now illustrate the above construction with an
example. The matrix $H$ below represents the directed graph $H$ that was
defined earlier. Thus its vertex set is the set of states observed in
the emitted sequence, which in our example is $\{1,2,\ldots , 14\}$. The
arcs of $H$ are labelled $h$ ({\em high}) or $l$ ({\em low}).

\vspace{0.1in}
\begin{center}
  $H = $
  \begin{tabular}{c|c|c|c|c|c|c|c|c|c|c|c|c|c|c}
    & 1 & 2 & 3 & 4 & 5 & 6 & 7 & 8 & 9 & 10 & 11 & 12 & 13 & 14 \\
    \hline 1 & $l$ & $h$ & $h$ & 0 & 0 & 0 & 0 & 0 & $l$ & $l$ & $h$ & $h$ & $h$ & $h$ \\
    \hline 2 & $l$ & $l$ & 0 & $h$ & 0 & 0 & $h$ & $h$ & $h$ & $h$ & 0 & $l$ & $h$ & $l$ \\
    \hline 3 & $l$ & 0 & $l$ & $h$ & $h$ & 0 & 0 & 0 & $h$ & $h$ & $l$ & $h$ & $h$ & 0 \\
    \hline 4 & $l$ & 0 & 0 & $l$ & 0 & $h$ & $h$ & $h$ & $h$ & $h$ & $h$ & 0 & $l$ & 0 \\
    \hline 5 & $l$ & 0 & 0 & 0 & $l$ & $h$ & $h$ & $h$ & $h$ & $h$ & $h$ & $l$ & $l$ & 0 \\
    \hline 6 & $l$ & 0 & 0 & 0 & 0 & $h$ & 0 & 0 & $l$ & $h$ & $h$ & 0 & $l$ & 0 \\
    \hline 7 & $l$ & 0 & 0 & 0 & 0 & 0 & $h$ & $h$ & $h$ & 0 & 0 & 0 & $l$ & 0 \\
    \hline 8 & $l$ & 0 & 0 & 0 & 0 & 0 & $h$ & $h$ & $h$ & 0 & 0 & 0 & $l$ & 0 \\
    \hline 9 & $l$ & $h$ & $h$ & 0 & $l$ & $h$ & $h$ & $h$ & $h$ & $h$ & $h$ & $h$ & $h$ & $h$ \\
    \hline 10 & $l$ & 0 & 0 & $l$ & 0 & $h$ & $h$ & $h$ & $h$ & $h$ & $h$ & 0 & $l$ & 0 \\
    \hline 11 & $l$ & 0 & $l$ & $h$ & $h$ & $h$ & 0 & 0 & $h$ & $h$ & $h$ & $h$ & $h$ & 0 \\
    \hline 12 & $l$ & $l$ & 0 & $h$ & $l$ & $h$ & $h$ & $h$ & $h$ & $h$ & $h$ & $l$ & $l$ & $l$ \\
    \hline 13 & $l$ & $h$ & $h$ & $h$ & $h$ & $h$ & $h$ & $h$ & $h$ & $h$ & $h$ & $h$ & $l$ & $h$ \\
    \hline 14 & $l$ & $l$ & 0 & $h$ & 0 & 0 & $h$ & $h$ & $h$ & $h$ & 0 & $l$ & $h$ & $l$ \\
  \end{tabular}
\end{center}
\vspace{0.1in}

Observe that the rows 6, 7 and 8 have the minimum number 3 of $h$,
therefore, $m = 3$, and $\cup_i \bar{U}_i = \{6,7,8\}$, where the union
is over the indices of the founders. Also, observe the block structure
of the sub-matrix consisting of rows and columns 6, 7 and 8: there are
no arcs from 6 to 7 or 8, and no arcs from 7 or 8 to 6, but there are
arcs between 7 and 8. Therefore, there are two founders in the pedigree.
There are outgoing arcs $(6,10)$ and $(6,11)$ that are labelled $h$,
therefore, the character set for one of the founders is $U_f =
\{6,10,11\}$. Similarly, the character set for the other founder is $U_g
= \{7,8,9\}$. We have called them $U_f$ and $U_g$ since we do not know
how many vertices are in the pedigree; but the naming is not relevant.
We now set $\mathcal{F} = \{\bar{U}_f = \{6\}, \bar{U}_g = \{7,8\}\}$.

We now consider pairs $\bar{U}_j$ and $\bar{U}_k$ in S. In this case
there is only one pair. The matrix $H$ shows 6 states 4,5,9,10,12,13
that have {\em high}-arcs to 6 and to $\{7,8\}$, and are therefore the
candidate states for inclusion in $\bar{U}_i$ for children $i$ of $j$
and $k$. We omit 10 from this list because 10 is in $U_f$ but not in
$\bar{U}_f$. We then note that $d^+_h(4) = d^+_h(5) = 6$, while
$d^+_h(9)$, $d^+_h(12)$, and $d^+_h(13)$ are all more than 6. Therefore,
we eliminate 9, 12 and 13 as well from the list of candidate states.
Since there are no arcs between 4 and 5, the blocks to be included in
$\mathcal{F}$ are $\bar{U}_e = \{4\}$ and $\bar{U}_d = \{5\}$. Both $d$ and $e$
are children of $f$ and $g$. Here we also conclude that since 9, 10, 11,
12 and 13 are in $U_d\cup U_e \cup U_f \cup U_g$, they cannot be in any
$\bar{U}_i$ that will be discovered in future, so they do not have to be
considered.

Next we have to repeat the process for all pairs of blocks in $\mathcal{F}$
(except of course the ones which we have already processed in earlier
steps).

Consider the pair $\bar{U}_e$ and $\bar{U}_g$. The states $2,12,13,14$
have {\em high}-arcs to each state in $\bar{U}_e \cup \bar{U}_g =
\{4,7,8\}$. But 12 and 13 have been eliminated before. Since $d^+_h(2) =
d^+_h(2) = 6$, and there are arcs $(2,14)$ and $(14,2)$, there is only
one new block $\bar{U}_c = \{2, 14\}$, and $c$ is a child of $e$ and
$g$.

Next we claim that $d$ and $g$ have no child together since only state
13 has {\em high}-arcs to all states in $\bar{U}_d\cup \bar{U}_g =
\{5,7,8\}$, but 13 has been eliminated earlier. By similar reasoning, we
claim that vertices $e$ and $f$ do not have a child, and vertices $d$
and $f$ do not have a child.

Next we note that the states 3, 11 and 13 have {\em high}-arcs to all
vertices in $\bar{U}_d\cup \bar{U}_e = \{4,5\}$. But 11 and 13 were
eliminated earlier. Therefore, the next block to be added to $\mathcal{F}$ is
$\bar{U}_b = \{3\}$.

Only 11 and 13 have {\em high}-arcs to all states in $\bar{U}_f$ and
$\bar{U}_d$. But 11 is in $U_f$, where $f$ is a founder, and 13 has {\em
high}-arcs to vertices in $\bar{U}_g$. Therefore, $d$ and $f$ have no
children together.

In the end, we observe that the states 1, 9, and 13 have {\em
high}-arcs to states in $\bar{U}_b \cup \bar{U}_c$, but 9 and 13 are
discarded before, so we conclude the construction by adding block
$\bar{U}_a = \{1\}$ to $\mathcal{F}$, which corresponds to the
extant vertex. The resulting pedigree is the one shown on the left
of Figure~\ref{fig-definition}.

\end{document}